\documentstyle[proceedings,psfig]{heliosymp} 

\newcommand{\la}{\mathrel{\hbox{\rlap{\hbox{\lower4pt\hbox{$\sim$}}}\hbox{$<$}}}}
\newcommand{\ga}{\mathrel{\hbox{\rlap{\hbox{\lower4pt\hbox{$\sim$}}}\hbox{$>$}}}}

\begin{opening}
\title{INTERNAL ROTATION, MIXING AND\protect\\  LITHIUM ABUNDANCES}

\author{Brian Chaboyer}
\institute{Steward Observatory\\
University of Arizona\\
Tucson, AZ, USA~~85710}

\end{opening}

\runningtitle{Rotation, Mixing and Lithium}

\begin{document}
{\fontsize{10pt}{11pt}\selectfont 
\begin{abstract}
Lithium is an excellent tracer of mixing in stars as it is destroyed
(by nuclear reactions) at a temperature around $\sim 2.5\times 10^6$
K.  The lithium destruction zone is typically located in the radiative
region of a star.  If the radiative regions are stable, the observed
surface value of lithium should remain constant with time.  However,
comparison of the meteoritic and photospheric Li abundances in the Sun
indicate that the surface abundance of Li in the Sun has been depleted
by more than two orders of magnitude.  This is not predicted by solar
models and is a long standing problem.  Observations of Li in open
clusters indicate that Li depletion is occurring on the main sequence.
Furthermore, there is now compelling observational evidence that a
spread of lithium abundances is present in nearly identical stars.
This suggests that some transport process is occurring in stellar
radiative regions.  Helioseismic inversions support this conclusion, for they
suggest that standard solar models need to be modified below the base
of the convection zone. There are a number of possible theoretical
explanations for this transport process. The relation between Li abundances,
rotation rates and the presence of a tidally locked companion along
with the observed internal rotation in the Sun indicate that the
mixing is most likely induced by rotation.  The current status of
non-standard (particularly rotational) stellar models which attempt to
account for the lithium observations are reviewed.

\end{abstract}

\section{Introduction}
Li\footnote{In this review I will use Li to represent $^7$Li, the
isotope which is produced by big bang nucleosynthesis.  $^7$Li
accounts for $\sim 93\%$ of the total Li abundance in meteorites.
Observers typically measure the total Li abundance, while
theoretical models determine depletion factors for $^7$Li and $^6$Li
The $^6$Li isotope is destroyed at much lower temperatures
and $^7$Li.  When making the comparison between the observations and
theory, it is usually assumed that the $^6$Li contribution to the
stellar Li content is neglible. However, see the discussion on halo
stars (\S \ref{secthalo}) where observations of $^6$Li may be used to
elucidate the mixing mechanism operating in these stars.}  is a
sensitive tracer of mixing in stellar radiative regions as it is easily
destroyed at temperatures above $\sim 2.5 \times 10^6\,$K.  For solar
type stars, the Li destruction region is located below the surface
convection zone in standard models.  As a consequence, standard
stellar models predict that Li should not be depleted at the surface
of solar type stars.  This is a rather robust prediction of stellar
evolution theory, which has been known for 40 years \cite{schwar}.
However, comparisons between the solar photospheric Li abundance and
the Li abundance in meteorites show that the Sun has depleted a
substantial amount of Li at its surface \cite{green51}. The solar Li
depletion problem has posed a challenge to stellar evolution theory
for 40 years, and the solution to this puzzle is still open to debate.

The Sun is unique in that helioseismic observations allow us to probe
the interior structure and rotation of the Sun.  These observations
can put constraints on possible solutions to the solar Li depletion
problem, but by themselves solar observations cannot uniquely
determine the cause of solar Li depletion.  Observations of stellar Li
abundances allow one to study the Li depletion problem as a function
of age, metallicity and stellar mass.  As such, they provide a
powerful test for mechanisms which attempt to explain the solar Li
depletion. The discovery of a large dip in Li abundances around 6600 K
in the Hyades \cite{lidip} was not predicted by theorists, and remains
a major challenge to theoretical stellar evolution models.  There is
increasing evidence that a dispersion in Li abundances exists among
stars with similar ages, metallicities and masses 
(\citeauthor{soder93} \citeyear{soder93}; \citeauthor{boes98}
\citeyear{boes98}).  Such a dispersion suggests that another stellar
property is important in determining the amount of Li which is
depleted in stars.  There is mounting observational evidence that
rotation plays a key role in determining the amount of Li depletion in
a star (\citeauthor{hyadbin} \citeyear{hyadbin}; \citeauthor{jones97}
\citeyear{jones97}).
In this review, I will discuss the relationship between mixing,
rotation and Li abundances in stars. 

\section{Solar Observations \label{sectsolar}}
The present photospheric abundance of Li in the Sun has been depleted
by a factor of $140\pm 40$ as compared to the meteoritic value of
$\log {\rm N(Li)} = 3.31\pm 0.04$ \cite{anders}\footnote{Using
the standard notation where $\log {\rm N(Li)} \equiv 12 + \log [{\rm
N(Li)/N(H)}]$.}.  Solar models which do not allow for transport or
mixing in the radiative regions of the Sun predict very little Li
depletion (a factor of 2 -- 3).  This has been a remarkably robust
prediction of standard stellar structure theory which has remained
unchanged for 40 years (\citeauthor{schwar} \citeyear{schwar};
\citeauthor{mysolar} \citeyear{mysolar}), despite the fact that the 
opacities and nuclear reaction rates used in stellar structure codes
have changed considerably over this time period.  Current solar models
imply that the region where Li is destroyed in the Sun is $\sim
0.05\,{\rm R}_\odot$ below the base of the solar convection zone.  The
Li depletion in the Sun suggests that some material from the region
where Li is destroyed has been transported to the convection zone,
leading to the observed Li depletion at the surface.  

The cause of this transport process is still a matter of debate.  
Observations of Be can constrain the nature of this transport process,
as Be is burned at a higher temperature ($3.5\times 10^6\,$K) than Li,
and so probes the deeper interior of the Sun.  Observations of the
photospheric and meteoritic Be abundances suggest that Be has been
depleted at the solar surface by a factor of $1.8\pm 0.5$ \cite{anders}.  
Thus, at the $3\,\sigma$ level it appears that the Sun has indeed
depleted Be at its surface.  This is supported by the work of
\citeauthor{solarbe} \shortcite{solarbe}, who found a minimum surface
solar Be  depletion level of 12\%.  

\begin{figure}
\centerline{\psfig{figure=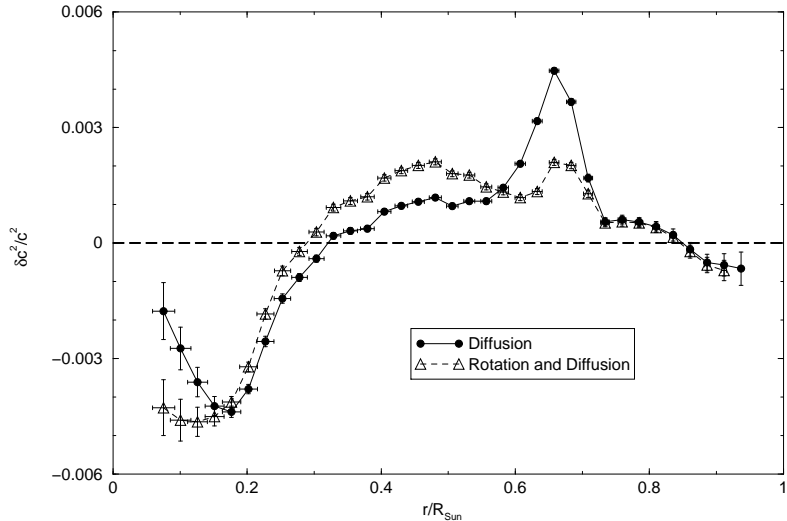,height=7.0cm,angle=270}  }
\caption{The difference in the square of the sound speed ($c^2$)
between two solar models and the actual Sun, as determined from
helioseismology.}
\label{figsound}
\end{figure}
Helioseismic observations can be used to probe the interior structure
and rotation of the Sun and provide a strong test for stellar
evolution theory.  The observed frequencies of the solar $p$-modes
depend primarily on the sound speed $c$ in the solar interior and so
it is relatively straightforward to invert the helioseismic
observations to probe the variation of sound speed with depth in the
Sun.  In order to linearize the problem, this inversion is typically
done with respect to a theoretical solar model.  The data which are now
available from the GONG project \cite{gong} and SOHO satellite
\cite{soho} allow for a very precise determination of the sound speed
in the Sun, down to $r \simeq 0.1\,{\rm R}_\odot$.  A typical example
of a solar sound speed inversion is shown in Figure \ref{figsound},
which shows that current solar models are able to match the observed
value of $c^2$ to within 0.5\% throughout the solar interior.  This is
a remarkably achievement for stellar evolution theory, made possible
by many advances in the last 5 years in the opacities and equation of
state used within stellar evolution codes.  Another important advance
has been the realization that diffusion (whereby the elements heavier
than hydrogen sink to the center of the star) must be included in
solar models in order to match the helioseismic observations
\cite{diff1,diff2}.

Although the match between the helioseismic observations and
theoretical models is impressive, the observations are extremely
accurate, and the remaining differences between the models and the Sun
can be used to continue to improve the physics used in stellar models.
Figure \ref{figsound} indicates that there is a large error in the
diffusion model near the base of the solar convection zone ($r \sim
0.7\,{\rm R}_\odot$).  Detailed studies of this region have suggested
that this error is due to the sharp change in the mean molecular
weight which occurs at the base of the solar convection zone in
models which include diffusion \cite{gongstruct}.  The most likely
explanation is that some form of slow turbulent mixing is operating
below the base of the solar convection zone \cite{basu}.  The
model labeled `rotation and diffusion' in Figure
\ref{figsound} includes slow turbulent mixing generated by rotation 
\cite{mysolar},
and provides a much better match to the observed Sun near the base of
the convection zone than the model which only includes diffusion.
Helioseismic observations provide strong evidence that slow
turbulent mixing occurs below the base of the convection zone in the
Sun.

Helioseismology can also probe the interior rotation rate of the Sun.
The global modes of oscillation in the Sun can be described by the
spherical harmonics, of radial order $n$, degree $\ell$, and
horizontal order $m$ (cf.\ \citeauthor{gongrot} \citeyear{gongrot}).
If the Sun were spherically symmetric, then the observed oscillations
would depend only on $n$ and $\ell$.  Rotation breaks the spherical
symmetry of the Sun, leading to a splitting of the common $n, \ell$
modes into different $m$ values.  The degree of this splitting can be
used to infer the solar rotation rate as a function of depth within
the Sun. Current data allows for a reliable rotation rate inversions
down to $r \sim 0.4\,{\rm R}_\odot$. Rotation rate rate inversions
done by a number of authors, using different data sets have all
reached essentially the same conclusion.  In the the radiative
interior there is quasi-rigid rotation, while a latitude-dependent
rotation exists in the entire convection zone
\cite{gongrot,lowl,sohorot}.  The transition between these two regimes
occurs in a thin layer below the base of the convection zone and is
referred to as the solar tachocline.  Recent work suggests that the
tachocline is very narrow, with a width estimated to be $r = (0.020\pm
0.005)\,{\rm R}_\odot$ \cite{basu} or $r = (0.05\pm 0.03)\,{\rm
R}_\odot$ \cite{tachocline}.  

\section{Young Cluster Observations \label{sectyoung}}
Observations of Li in young cluster stars can be used to empirically
determine the amount of stellar Li depletion as a function of age,
mass and chemical composition.  These observations provide very strong
constraints for stellar models.  A striking result from early
observations of Li abundances in cluster stars was the existence of a
dip in Li abundances in the F stars (around ${\rm T_{eff}} \simeq
6600\,$K) in the Hyades cluster \cite{lidip}.
Stars around this temperature range have Li abundances which are at
least 1.0 dex lower than stars which are hotter or cooler than the dip
(see Figure \ref{fighyad}).  This dip was not predicted by theoretical
models.  Stars on the hot side of the
Li dip do not have convective envelopes, and stars on the cool side of
the Li dip have small convective envelopes.  The Li
dip occurs in  stars with very small convective envelopes. The Hyades is
a somewhat metal-rich cluster (${\rm [Fe/H]} = +0.1$, \cite{hyadfeh})
which is approximately 600 Myr old \cite{hyadhipp}.  Observations of
Li abundances in other clusters (the Pleiades in particular) have
shown that the Li dip does not exist for stars on the zero age main
sequence (ZAMS)
\cite{soder93}. 
\begin{figure}
\centerline{\psfig{figure=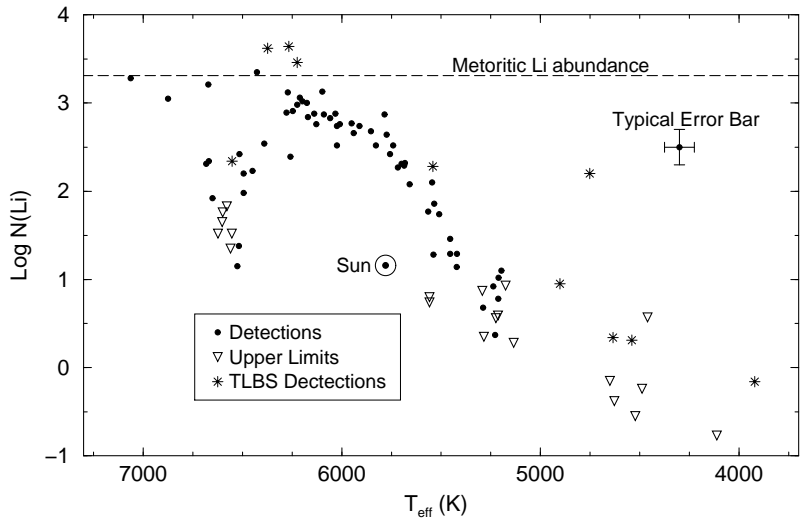,height=7.0cm,angle=270}  }
\caption{Observations of Li in the Hyades
(Balachandran 1995; Barrado y Navascu\'{e}s \&  Stauffer 1997).
The photospheric solar Li abundance is indicated by the solar symbol
$\odot$, and the meteoritic Li abundance is given by the dashed line.}
\label{fighyad}
\end{figure}

In addition to the Li dip, there are a number of interesting features
in Figure \ref{fighyad}.  In particular, it appears that the tidally
locked binary stars (TLBS) have (on average) a higher Li abundance
than single stars or binaries which are not tidally locked
\cite{hyadbin}.  This is in good agreement with the theoretical work
of \citeauthor{zahn} \shortcite{zahn}, who postulated that Li
depletion is due to rotational mixing and this rotational mixing does
not operate in TLBS.  The excess Li abundance observed in TLBS is
strong evidence that rotation plays a key role in Li depletion.

The Hyades stars around the solar temperature (${\rm T_{eff}} =
5780\,$K) have significantly higher Li abundances in the Sun (Figure
\ref{fighyad}).  This is despite the fact that the Hyades is slightly
more metal-rich than the Sun and so theoretical models would predict
that the Hyades stars should deplete more Li than the Sun on the
pre-main sequence.  This suggests that Li depletion in these cooler
stars (${\rm T_{eff}} \la 6000\,$K) occurs on the main sequence, a
conclusion which is reinforced by the observed Li abundances in the
Pleiades \cite{myopen}.

Another interesting feature of Figure \ref{fighyad} is that the
highest Li abundances observed are similar to the meteoritic Li
abundances, suggesting that their has been no significant Li
enhancement over the last 4 Gyr.  It is important to note
that the only stars which have Li abundances similar to the meteoritic
abundances are either TLBS, or are on the hot side of the dip (and hence,
do not have a surface convection zone).  The question then arises, do
all single stars with surface convection zones deplete Li?  The stars
which have the least amount of Li depletion are those on the cool side
of the Li dip, around ${\rm T_{eff}} = 6200\,$K.  In an seminal paper,
\citeauthor{boes91} \shortcite{boes91} demonstrated that, even for the
stars around ${\rm T_{eff}} = 6200\,$K, there was a clear correlation
between their Li abundance and age.  Older stars have lower Li
abundances, implying that all stars with surface convection zones
deplete Li on the main sequence.  The work of \citeauthor{boes91}
\shortcite{boes91}  utilized observations
from a number of different observers, and included stars with a rather
large range in effective temperatures.  In order to see if this had
any effect on her conclusions, I have repeated the analysis of
\citeauthor{boes91} \shortcite{boes91} using the uniform data set
compiled by \citeauthor{bala} \shortcite{bala} and only included stars
with $6300 \le {\rm T_{eff}} \le 6100\,$K.  The results are shown in
Figure \ref{figdeplet}.  Even with this restricted data set it is
clear that older stars have lower Li abundances, implying that Li
depletion has occurred on the main sequence for stars around ${\rm
T_{eff}} = 6200\,$K. It is interesting to note that halo stars in the
same temperature range appear to match the trend found in open cluster
stars (see \S \ref{secthalo}).
\begin{figure}
\centerline{\psfig{figure=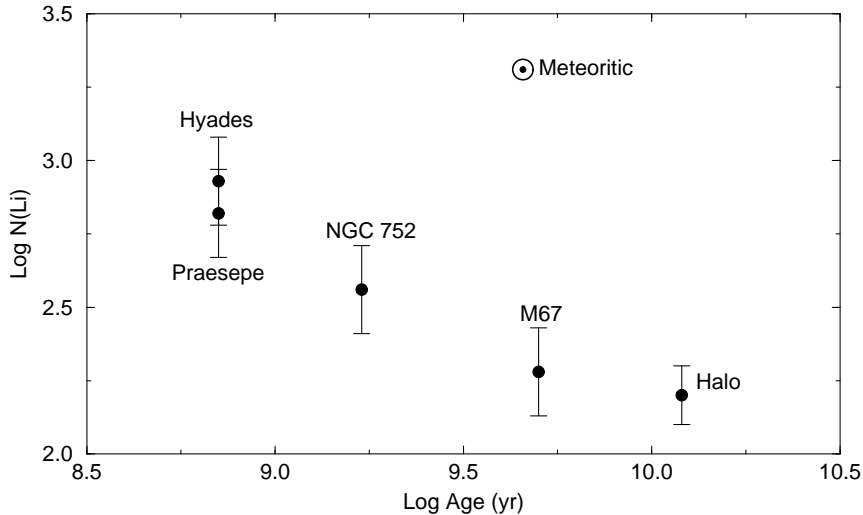,height=7.0cm,angle=270}  }
\caption{The mean abundance of Li for stars in the temperature range 
$6300 \le {\rm T_{eff}} \le 6100\,$K in four different
clusters, as a function of the cluster age (data from Balachandran
1995).  The mean abundances for stars in the same temperature range in
the halo is taken from the data of Ryan {\it et al.}\ (1996).
The meteoritic Li abundance is also indicated.}
\label{figdeplet}
\end{figure}

The existence of a dispersion in Li abundances at a given effective
temperature within a cluster has been the subject of a number of
observational papers.  If such a dispersion exists, it proves that
something besides age, mass and chemical composition must control Li
depletion in stars.  \citeauthor{thorburn} \shortcite{thorburn}
studied in detail the existence of a dispersion among Hyades stars.
They found that a such a dispersion does exist for stars with ${\rm
T_{eff}} = 6100\,$K, with $\sigma_{\rm N(Li)} \sim 0.15\,$dex.  The
existence of a dispersion among the cooler stars was more difficult to
prove, though the data did suggest that a small dispersion of 
$\sigma_{\rm N(Li)} \sim 0.09\,$dex existed among the G stars
(${\rm T_{eff}} \sim 5500\,$K).  In contrast, there is clear evidence
for a Li dispersion in this temperature range among the
Pleiades (a solar metallicity cluster whose stars are on the ZAMS)
stars, while there is little evidence for a dispersion among
the hotter Pleiades stars  \cite{soder93}.   There was also clear
evidence for a correlation between rotation rates and Li abundances
in the Pleiades, with the fastest rotators having the highest Li
abundances.  A recent paper by \citeauthor{jones97} \shortcite{jones97}
studies in detail the dispersion in Li abundances (at a given
effective temperature) among the Pleiades (age 70 Myr), M34 (age 250
Myr) and the Hyades (age 600 Myr). They found that the dispersion
among the lower mass stars (${\rm M} \sim 0.65$ to $0.95\,{\rm
M}_\odot$) is greatest on the ZAMS, and decreases with age.  The
fastest rotating stars have the highest Li abundances.  In
addition, they found that Li was depleted on the main sequence,
leading \citeauthor{jones97} to conclude ``{\it high rotation
preserves lithium during pre-main sequence evolution and that a high
angular momentum loss rate accelerates lithium depletion after the
star is on the main sequence\/}''.  This is a key observational fact,
which implies that rotation must lead to mixing and Li depletion in
low mass main sequence stars. 

However, the basic assumption used in the above analysis, that a
dispersion in Li equivalent widths corresponds to a dispersion in Li
abundances has been brought into question by \citeauthor{stuik}
\shortcite{stuik}.   These authors suggest that magnetic activity
(which leads to the formation of pots and plage on cool stars) may
effect the Li line strength.  Thus, the observation that fast
rotators have high Li equivalent widths could be due to the fact that
fast rotators are more active stars than slow rotators.  The work of 
\citeauthor{stuik} \shortcite{stuik} is based upon theoretical models
(which have known problems), and is a schematic feasibility analysis,
rather than a definitive statement that magnetic activity
effects Li line strengths.  Nevertheless, it raises a key point which
clearly requires more attention in the future.

\section{Halo Star Observations \label{secthalo}}
Observations of Li in very old stars has the potential to determine
the primordial Li abundance.  Li is one of only 3 elements (the others
being H and He) which are believed to have been produced in significant
amounts by big bang nucleosynthesis.  Thus, the determination of the
primordial Li abundance is a key constraint on theoretical models of
big bang nucleosynthesis.  In their pioneering work, \citeauthor{spite}
\shortcite{spite} measured Li abundances in 13 metal-poor (halo) stars, and
found that all of the stars had nearly identical Li abundances.  As
there was no relation between Li abundances and metallicity, this
suggested that the observed Li abundance was the primordial Li
abundance.  Since then, there have been numerous studies of Li
abundances in halo stars.  These studies confirmed that halo
stars with ${\rm T_{eff}} \ga 5600\,$K have nearly identical Li
abundances, independent of their metallicity or temperature 
\cite{sms,rmb,hd1,ht1}.  

However, this assumption was brought into question by the discovery of
a number of hot halo stars which have abundances significantly lower
than the plateau \cite{hwt,spite93,tb93a,t92}. About $\sim 5\%$ of the
halo stars in the plateau effective temperature range have low Li
abundances.  These stars demonstrate that at least some hot halo stars
do deplete Li, leaving open the possibility that plateau stars may
have also depleted Li.  The current status of Li abundances
measurements in halo stars is shown in Figure \ref{fighalo}.  
\begin{figure}
\centerline{\psfig{figure=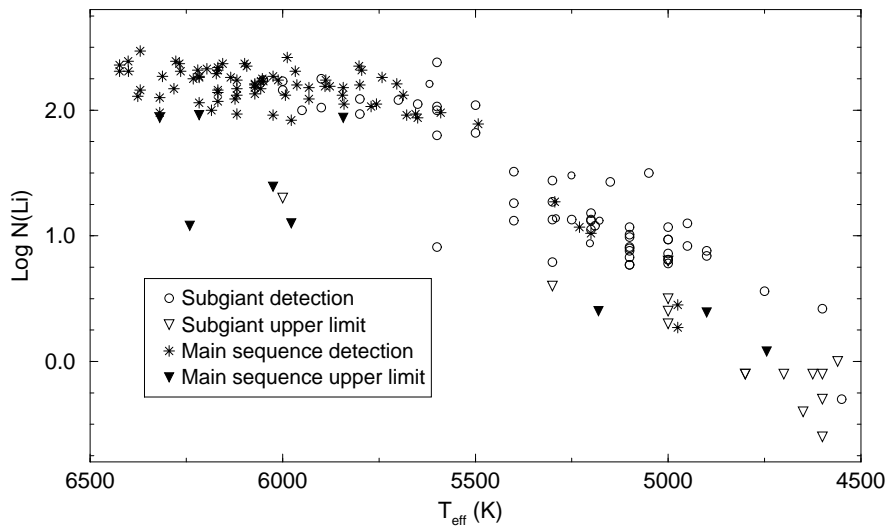,height=7.0cm,angle=270}  }
\caption{Li abundances in low metallicity (${\rm [Fe/H]} < -1.0$)
field stars (halo stars).  Data from Pilawchoski {\it et al.}\ (1993),
Thorburn (1994),  and Deliyannis {\it et al.}\ (1994).}
\label{fighalo}
\end{figure}

The existence of a dispersion among the Li plateau stars would
strongly suggest that even these stars have depleted Li.  The question
of weather or not a dispersion exists among the plateau stars is an
open one.  \citeauthor{dpd93} \shortcite{dpd93} performed a dispersion
analysis in the Li equivalent width--colour data and concluded that an
intrinsic Li/H dispersion of 10\% existed among the plateau
stars. Furthermore, a uniform analysis of $\sim 80$ plateau halo stars
found a correlation between Li abundances, effective temperatures and
metallicity \cite{thorhalo}.  Hotter, and/or more metal-rich stars
were found to have higher Li abundances.  This work was confirmed by 
\citeauthor{ryan} \shortcite{ryan}.  Using a sub-set of the above
observations, and adopting a different effective temperature scale 
\citeauthor{molaro} \shortcite{molaro} and \citeauthor{bonifacio} 
\shortcite{bonifacio} found no evidence for a dispersion, or
correlation with [Fe/H]. 

In order to directly probe for the existence of a Li dispersion among
hot halo stars, \citeauthor{boes98} \shortcite{boes98} have
determined Li abundances of 7 stars in the globular cluster M92 (${\rm
[Fe/H]} = -2.1$).  Stars in a given globular cluster all have the same
metallicity, age and reddening.  In addition, it is possible to
observe a number of stars with virtually the same colour, implying
that the stars all have the same effective temperature.
\citeauthor{boes98} \shortcite{boes98} observed 4 stars with virtually
identical effective temperatures, of which 1 star has a significantly
higher Li abundance than the others.  In all other aspects (including
a detailed element by element analysis), the star with the high Li
abundance is virtually identical to the other 3 stars.  This
observation strongly suggests that a dispersion in Li abundances
exists among hot halo  stars and that this dispersion is due to
stellar depletion from a higher primordial Li abundance.  

Observations of $^6$Li can also be used to determine if the plateau stars
have depleted Li.  $^6$Li is a destroyed at much lower temperatures
than $^7$Li, so that if enough mixing has occurred to deplete $^7$Li,
virtually all of the $^6$Li should be destroyed.  An observation of
$^6$Li in a hot halo star would suggest that there has not been
any significant depletion of $^7$Li in these stars, and that the
observed $^7$Li abundance is very close to the primordial
abundance\footnote{Standard big bang nucleosynthesis predicts that all
of the Li production will be in the form of $^7$Li.  $^6$Li can be
produced by cosmic ray spallation in the interstellar medium.  
Deliyannis  \& Malaney (1995) have suggested that 
$^6$Li may be produced by stellar flares at the surface of the star,
implying that an observation of $^6$Li in the photosphere of a
star does not constrain the depletion of $^7$Li in the stellar
interior.}.  The best evidence for 
a $^6$Li detection in a hot halo star has been presented by 
\citeauthor{smith93} \shortcite{smith93} and \citeauthor{ht97}
\shortcite{ht97}.  Both of these groups claim a detection of $^6$Li in
the subgiant HD 84937 (${\rm [Fe/H]} = -2.4$, ${\rm T_{eff}} =
6100\,$K).  The careful analysis of \citeauthor{ht97} \shortcite{ht97}
led them to conclude that $^6{\rm Li}/^7{\rm Li} = 0.08\pm 0.04$ in HD
84937.  Thus, the detection is significant at the $2\,\sigma$ level.
It is premature to conclude that $^6$Li has definitively been detected
in a hot halo star.  

Halo stars cooler than the plateau show progressively lower lithium
abundances, similar to that seen in open clusters.  This is true both
for main sequence stars, and subgiant stars (see Figure
\ref{fighalo}).  In both cases, the deepening convective envelopes
have brought Li depleted material to the surface of the star.  It is
interesting to note that stellar evolution models which incorporate
the latest available input physics are unable to correctly reproduce
the onset of Li depletion observed in main sequence or subgiant halo
stars \cite{myhalo}.  This implies that the models are in need of
revision, and casts doubt on the ability of the models to correctly
predict the amount of $^6$Li depletion which occurs.

\section{Theoretical Models \label{secttheory}}
The preceding sections have made it clear that Li depletion has
occurred in stars with a variety of masses and ages, for which standard
stellar evolution models predict no Li depletion.  A number of
mechanisms have been proposed which deplete Li, in order to explain
some or all of the Li observations.  These Li depletion mechanisms
include: overshoot, mass loss, mixing induced by gravity waves, 
diffusion, and mixing induced by rotation.

Overshoot at the base of the convection zone has been proposed by a
number of authors (most recently by \citeauthor{ahrens}
\citeyear{ahrens}) to explain the depletion of Li in the Sun.  Solar
models which include an overshoot of $\sim 0.24$ pressure scale
heights at the base of the convection zone are able to reproduce the
observed solar Li depletion.  However, this Li depletion occurs on
the pre-main sequence, implying very low Li abundances in young, cool (${\rm
T_{eff}} \sim 5800\,$K) stars.  This is not seen in any of the
open cluster observations (including the Hyades, Figure \ref{fighyad}
and the Pleiades).  Open cluster observations rule out fast overshoot
at the base of surface convection zones as a viable explanation of the
observed solar Li depletion \cite{myopen}.  

Large amounts of mass loss ($\sim 0.05\,{\rm M}_\odot$) sufficient to
expose Li depleted material at the surface of a star has been suggested as a
cause of the F star Li dip \cite{schramm}.  However, a detailed stellar
evolution study by \citeauthor{swen} \shortcite{swen} found that mass
loss alone could not explain all open cluster Li observations.  
Furthermore, a detailed study of Be abundances among Li dip stars has
 detected moderate Be deficiencies among stars with severe
(but detected) Li abundances \cite{beli}.  The mass loss hypothesis
requires that all of the Li be depleted before any Be depletion
occurs, in contradiction with these observations.

\citeauthor{press} \shortcite{press} suggested that internal
gravity waves generated at the base of the convective envelope of a
star may produce weak mixing in the radiative interior.  This
suggestion was studied in detail by \citeauthor{lopez}
\shortcite{lopez}, who found that mixing induced by internal gravity
waves could explain the Li dip observed in F stars if the intensity of
the gravity waves was increased by a factor of 15 above mixing length
estimates.  The degree of Be depletion predicted by these observations
appears to be incompatible with the observations of \citeauthor{beli}
\shortcite{beli}.

The diffusion of Li out of the surface convection zone was suggested
by \citeauthor{michaud} \shortcite{michaud} to be the cause of the Li
gap.  The models required a small mass loss rate ($10^{-15}\,{\rm
M_\odot\,yr^{-1}}$) to explain the lack of Li depletion on the hot side
of the dip. Helioseismology indicates that He diffusion occurs in the
Sun (\S \ref{sectsolar}), indicating that Li diffusion should also be
operating in stars.  As the  diffusion time scales for Li and Be are
similar, Be and Li should be depleted by the same amount if 
diffusion is the cause of the Li dip.
Observations indicate that the degree of Be depletion is much smaller
than Li depletion in dip stars \cite{beli}, ruling out diffusion as
the sole cause of the Li dip.  Diffusion is also unable to explain the
Li depletion observed in the cooler stars (${\rm T_{eff}} \la
6000\,$K) like the Sun \cite{myopen}.

It has been known for over 70 years that thermal imbalances in
rotating stars give rise to large scale flows
\cite{vonzeipel,eddington,sweet}.  Thus, it is not surprising that a
number of authors have suggested that the Li observations can be best
explained by mixing induced by rotation 
(e.g.\ \citeauthor{char88} \citeyear{char88}; 
\citeauthor{vauclair} \citeyear{vauclair}; 
\citeauthor{pin90} \citeyear{pin90},\citeyear{pin92}; 
\citeauthor{char92} \citeyear{char92},\citeyear{char94};
\citeauthor{zahn92} \citeyear{zahn92};
\citeauthor{myhalo2} \citeyear{myhalo2};
\citeauthor{myopen} \citeyear{mysolar},\citeyear{myopen};    
\citeauthor{del97} \citeyear{del97}).
The exact physics underlying rotation induced mixing is still not well
understood, leading to a variety of approaches in dealing with the
mixing.  The Yale rotational models 
(\citeauthor{pin90} \citeyear{pin90},\citeyear{pin92};
\citeauthor{myhalo2} \citeyear{myhalo2};
\citeauthor{myopen} \citeyear{mysolar},\citeyear{myopen};
\citeauthor{del97} \citeyear{del97}) are perhaps the most ambitious, as
they self-consistently include the structural effects of rotation
along with the coupled transport of angular momentum and material due
to a variety of rotation induced mixing mechanisms.
These models, along with those of \citeauthor{char94}
\citeyear{char94} are able to reproduce a number of features of the
observed Li abundances, including the cool side of the Li dip and the
main sequence depletion of Li in the Sun and similar temperature
cluster stars.  Generically, these models also predict a correlation
between rotation velocities and Li abundances (observed in
young clusters) and that tidally locked binary stars should not
deplete Li \cite{zahn} as observed in the Hyades \cite{hyadbin}.
The rotation models do a good job of matching the Be observations
\cite{beli} in the Li dip stars \cite{del97}.  These models also
predict that a dispersion of Li abundances (due to a dispersion in
initial rotation velocities) should exist for stars with equal ages,
masses and metallicities, as observed in open clusters.  

The small (or non-existent) dispersion among the plateau stars in the
halo puts strong constraints on the rotation induced mixing models
\cite{myhalo2}, but does not rule them out.  Rotation induced mixing
models predict some main sequence Li depletion for all stars with
surface convection zones, in good agreement with the observations.  By
extension, these models also predict significant Li depletion among
metal-poor halo stars implying that the primordial abundance is higher than
the plateau value \cite{pin92,myhalo2}.  The exact amount of Li
depletion depends on the details of the models.  The Yale models
predict a factor of $\sim 10$ depletion, implying a primordial Li
abundance of $\log\,{\rm N(Li)} = 3.1$, a value which is incompatible
with standard big bang nucleosynthesis and current estimates for the
primordial abundances of helium and deuterium.  However, these models
do not match all of the observations, and there are considerable
uncertainties associated with these models.  Thus, the exact amount of
Li depletion which has occurred in the plateau stars is still an open
problem.

One of the key problems with the Yale rotational models is that
angular momentum transport was inefficient, leading to fast rotating
cores.  This is clearly in contradiction with helioseismic
observations of the solar internal rotation (\S \ref{sectsolar}).  An
easy way to correct this problem has recently been identified by 
\citeauthor{kumar} \shortcite{kumar} and \citeauthor{zahn97} 
\shortcite{zahn97}.  These authors found that 
low-frequency gravity waves could be excited by convection in the Sun.
These gravity waves can transport angular momentum very efficiently.
The estimates for the time scale of mixing by these two groups are very
similar to each other, and suggest that rigid body rotation would be
enforced in the solar radiative region on time scales of $10^7$ to
$10^8$ years.  Another difficulty with the Yale models which included
the  combined effects of rotation and diffusion was that the inclusion
of diffusion lead to a differential Li depletion across the effective
temperature of the halo Li plateau \cite{myhalo2}.  The derived
Li abundances did not agree with observations of the most metal-poor
stars.  This difficulty in reproducing the plateau Li abundances when
diffusion is included can be overcome by the addition of a modest 
stellar wind ($\sim 10^{-12.5}\,{\rm M_\odot\,yr}^{-1}$) in the models
\cite{vaucl95}.

\section{Summary\label{sectsumm}}
There is a wealth of data on Li abundances and rotation velocities in
stars with a variety of ages, masses and metallicities.  This data
clearly indicates that Li depletion occurs on the main sequence for
all stars with a surface convection zone.  This is in direct
contradiction with standard stellar evolution theory.
A number of possible mechanisms which lead to extra Li depletion have
been put forth. The dispersion in Li abundances at a given age,
metallicity and temperature, the 
correlation between Li abundances and rotation
velocities in Pleiades, the fact that tidally locked binary
stars in the Hyades have an excess Li abundance as compared to single
stars, and the detection of moderate Be deficiencies among Li dip stars
with detectable Li abundances, 
all imply that that rotation induced mixing is leading to Li
depletion on the main sequence.  Helioseismic observations of the Sun
support this hypothesis, for they show that slow form of slow mixing
is operating below the base of the solar convection zone \cite{basu}.
Current stellar models which incorporate rotation
induced mixing explain many, by not all of the observations.  Models
which are able to account for all of the data are likely to include
diffusion, rotation induced mixing, angular momentum transport by
gravity waves and/or magnetic fields and modest stellar winds.

} 
\end{document}